\documentclass[aps,prc,superscriptaddress,nofootinbib]{revtex4}
\usepackage{graphicx}

\begin{document}

\title{Investigation of collective radial expansion and stopping\\in heavy ion collisions at Fermi energies}

\author{E. Bonnet}
\email{eric.bonnet@ganil.fr}
\affiliation{GANIL, CEA-DSM/CNRS-IN2P3, Bvd. Henri Becquerel, F-14076 Caen CEDEX, France}
\author{M. Colonna}
\affiliation{INFN-Laboratori Nazionali del Sud, I-95123 Catania, Italy}
\author{A. Chbihi}
\author{J.D. Frankland}
\author{D. Gruyer}
\author{J.P. Wieleczko}
\affiliation{GANIL, CEA-DSM/CNRS-IN2P3, Bvd. Henri Becquerel, F-14076 Caen CEDEX, France}

\date{\today}

\begin{abstract}
We present an analysis of multifragmentation events observed in central Xe+Sn reactions
at Fermi energies. Performing a comparison between the predictions of the 
Stochastic Mean Field (SMF) transport model and experimental data,
we investigate the impact of the compression-expansion dynamics 
on the properties of the final reaction products.
We show that the amount of radial collective expansion, which characterizes the dynamical stage of the reaction,
influences directly the onset of multifragmentation and the kinematic properties of multifragmentation events.
For the same set of events we also undertake a shape analysis in momentum space, looking at
the degree of stopping reached in the collision, as proposed  in 
recent experimental studies. We show that full stopping is achieved for the most 
central collisions at Fermi energies.
However, considering the same central event selection as in the experimental data,
we observe a similar behavior of the stopping power with the beam energy, which
can be associated with a change of the fragmentation mechanism, 
from statistical to prompt fragment emission.
\end{abstract} 

\maketitle

\section{Introduction}

Nuclear matter Equation of State (EoS) and transport properties play a fundamental role in 
the understanding of many aspects of nuclear physics and astrophysics. For instance, neutron star genesis and cooling,
as well as the possible occurrence of hybrid structures, transition to the deconfined phase and black-hole formation,
are strongly influenced by nuclear compressibility and viscosity~\cite{astro}. 
Heavy ion collisions (HIC) offer the unique possibility to create, in terrestrial laboratories, 
transient states of nuclear matter in a wide range of density, temperature and N/Z asymmetry, 
thus making it possible to access information on fundamental properties of nuclear systems 
far from normal conditions.
Indeed relevant indications on the stiffness of the EoS at supra-saturation densities and on 
in-medium modifications of the two-body nucleon-nucleon cross section have emerged from the study 
of collective flows and degree of stopping measured in heavy ion reactions at 
relativistic energies~\cite{science}.
It is worth mentioning that collective flows are also widely investigated in the context of ultra-relativistic HIC, 
in connection with the occurrence of phase transitions to the deconfined Quark-Gluon-Plasma phase and 
its viscosity~\cite{enzo}.

Generally speaking, reaction mechanisms in the Fermi energy domain (20-50 MeV/nucleon) reflect an interesting interplay between 
mean-field (one-body) properties, dominant at low energies, and the increasing importance of two-body correlations such as nucleon-nucleon collisions.
In particular, the combined action of nuclear compressibility and nucleon-nucleon cross section governs 
the compression-expansion dynamics typically observed in central collisions. The reaction path can then bring 
the formed excited nuclear system to low density regions where the mean field becomes unstable. In this case two-body correlations and fluctuations are expected to provide the seeds
for fragment formation leading to the occurrence of multifragmentation phenomena
which can be described in the framework of a liquid-gas-type phase transition~\cite{rep}.

Observed first by the FOPI collaboration~\cite{Jeong,Hsi,Poggi,Kunde} and interpreted as an "extra push" with respect to the thermal pressure of  an equilibrated composite system, 
a collective expansion energy has been extracted from experimental data over a wide range of beam energy~\cite{mf}.
Several studies of the multifragmentation process, and of the corresponding role of the collective expansion, have been
undertaken in the framework of transport theories~\cite{rep,Li,Hagel,Chikazumi} and some analyses~\cite{Pal,Lopez,Gulminelli} have pointed out the importance 
of this effect on fragment formation, looking at, for example, the balance between the amount of radial collective flow and recombination probability~\cite{Pal}.

However, to our knowledge, estimates of the radial collective energy present in experimental multifragmentation
data have been mostly obtained by employing statistical models~\cite{MMM,MMMC,SMM} which treat separately
fragment production and collective expansion effects~\cite{Marie,Hudan,Bellaize,Lefevre,Piantelli,Beaulieu}. 
The main justification is the small contribution of the collective expansion energy~\cite{LG} with respect to the total excitation 
energy characterizing the Fermi energy domain (around 20-30\%). Nevertheless, even for this amount of expansion, recent 
experimental analyses 
show a significant influence on the fragment production~\cite{Bonnet}.

Particularly sensitive to the interplay between one- and two-body effects 
are also observables characterizing the shape of the reaction events in momentum space. 
For instance, the so-called stopping power, which was first investigated by the FOPI collaboration at higher beam energies~\cite{Reisdorf}, measures 
the efficiency of conversion of the initial beam energy into transverse
directions, as quantified by taking the ratio of total transverse to parallel
energy-based quantities. In Ref.~\cite{Lehaut}, similar analysis has been done for collisions at Fermi energy domain and one result is
that full stopping is not achieved.

The aim of this paper is to bring information on the connection between fragmentation features and 
the underlying collective expansion mechanism, with the help of a transport theory description of the reaction dynamics. 
Moreover, we will discuss the existence of possible correlations between the evolution of the fragmentation 
regime and the experimentally-observed trend for the amount of stopping reached in heavy ion reactions at Fermi energies.
A combined study of the two features appears as a promising tool to learn about relevant nuclear matter properties.
The reaction dynamics is described in the framework of a semi-classical microscopic transport model, 
the Stochastic Mean Field (SMF) approach~\cite{colonna98,colonna2004}. This choice fits the requirement
to have a well implemented nuclear mean-field dynamics together with the effects of fluctuations induced by two-body scatterings.
Central and mid-central collisions will be investigated taking, as reference, the INDRA data~\cite{Hudan, Marie, Rivet,Tabacaru} 
for the reactions $^{129}$Xe+$^{119}$Sn at 25,32,39,45 and 50 MeV/nucleon.

The paper is organized as follows: in Section II we first review the main ingredients of the SMF
model, then we briefly present the systems analysed before discussing SMF results concerning 
phase diagram trajectories, pre-fragment formation and recombination.
Section III is devoted to the comparison with INDRA data : for this purpose we first test, on simulated events, the 
validity of the adopted central collisions selection,
based on the flow angle observable~\cite{Marie}. Treating in the same way simulations and experimental data, we show a comparison of fragment partition properties
looking at charge and velocities. 
Observables related to stopping power in nuclear collisions are introduced and discussed in Section IV.
Finally, conclusions and perspectives are drawn in Section V. 

\section{Results of the SMF model}

\subsection{Description and ingredients}
We consider, as starting point, the Boltzmann-Langevin (BL) equation for the time evolution of
the semiclassical one-body distribution function $f({\bf r},{\bf p},t)$:
\begin{equation}
\frac{\partial f}{\partial t}+\frac{\bf p}{m}\frac{\partial f}{\partial {\bf r}}-
\frac{\partial U}{\partial {\bf r}}\frac{\partial f}{\partial {\bf p}}=I_{coll}[f]+\delta I[f].
\end{equation}
The coordinates of isospin are not shown for brevity.
Eq.(1) essentially describes the evolution of the system
in response to the action of the self-consistent mean-field potential $U$, whereas effects of two-body 
correlations and fluctuations are incorporated in the collision integral, $I_{coll}$, and its stochastic
part, $\delta I$.
The average term $I_{coll}[f]$ takes into account the 
energy, angular and isospin dependence of free nucleon-nucleon cross sections \cite{Baran2002}.

The SMF model represents an approximate approach to solve the BL equation, 
where phase-space fluctuations are projected in coordinate space. 
Thus the fluctuating term $\delta I[f]$ is implemented through stochastic spatial density fluctuations~\cite{colonna98,colonna2004}. 

We adopt the following parametrization of the mean-field potential:
\begin{eqnarray}
U_{q}&=&A\frac{\rho}{\rho_0}+B(\frac{\rho}{\rho_0})^{\alpha+1}
+C \frac{\rho_n-\rho_p}{\rho_0}\tau_q \nonumber \\
\end{eqnarray}
where $\rho$ denotes the density, $q=n,p$ and $\tau_n=1, \tau_p=-1$.
The coefficients $A=-356~MeV$,$B=303 ~MeV$ and the exponent $\alpha=\frac{1}{6}$, 
characterizing the isoscalar part of the mean-field, are fixed requiring 
that the saturation properties of symmetric 
nuclear matter, $\rho_0=0.16 fm^{-3}$ and $E/A=-16~MeV/nucleon$,
with a compressibility of $200$$ MeV$, are reproduced. This choice corresponds
to a Skyrme-like effective interaction, namely $SKM^*$, for which we consider the effective
mass as being equal to the nucleon bare mass. 
As far as the isovector part of the nuclear interaction is concerned, 
we take a constant value of $C = 36~MeV$, corresponding to a linear (stiff)
behavior of the potential part of the symmetry energy, 
$C_{sym}^{pot} = 36\rho/(2\rho_0)$~\cite{Baran2002,rep_isospin}.

Eq.(1) is solved adopting the test particle method. 
The inclusion of fluctuations in the dynamics allows one to address mechanisms governed by the growth
of mean-field instabilities, such as spinodal decomposition, which 
lead to multifragmentation process~\cite{rep}.
The products generated by the reaction dynamics are reconstructed from the one-body density distribution by applying a coalescence
procedure, 
connecting nearby cells with density larger than a cut-off value. 
At the end of the dynamical stage (i.e. at the so-called freeze-out time) 
the primary fragments are still hot. Hence the dynamical events are plugged in 
a statistical de-excitation code, in order to get the properties of the final reaction products. 
We adopt the SIMON code~\cite{Simon},
which describes the in-flight secondary de-excitation through the Coulomb field. 
A drawback of our mean-field based approach is that the production of light charge particles ($Z<3$, $A>1$) is not well described. Indeed the primary yield of such particles is largely underestimated in the code, thus favouring free nucleon emission. In the following we will concentrate on the properties
of intermediate mass fragments (IMF) with charge (Z) greater than 4. 

\subsection{Details of the calculations}

We performed, for the system $^{129}$Xe+$^{119}$Sn, two runs of calculation :
\begin{itemize}
\item The first one is dedicated to central collisions. The range of impact parameter (b) is [0;4] fm by step of 0.5 fm, 
the beam energies are 25,32,35,39,45, 50 MeV/nucleon. For the lowest beam energies (25 and 32 MeV/nucleon), 
the possible contribution of larger impact parameters, up to $b=6.5~fm$, is considered.
We use 30 test-particles per nucleon and 
run 1000 events for each system in a 40$\times$40$\times$40 $fm^{3}$ box until t=400 fm/c.
\item The second run is dedicated to the investigation of the stopping power. We take the same beam energies and enlarge 
the impact parameter range, b$\in$[0.5;6.5] fm. We run 100 events for each system in a 80$\times$80$\times$80 $fm^{3}$ box until t=440 fm/c.
\end{itemize}

We consider two values for the density cut-off employed in the coalescence procedure: 
0.03 and 0.09 $fm^{-3}$. The first value corresponds to the standard parameter used
in the model \cite{Baran2002}. According to this choice, all regions with density lower than 0.03 $fm^{-3}$ (about 1/5 of the 
saturation density) are associated with free nucleon emission. 
The large cut-off density of 0.09 $fm^{-3}$ is used for exploratory purposes. Indeed it allows one
to identify high-density peaks (a kind of early recognition of fragments) even at early times, when the
system is still rather compact, and 
to investigate the recombination effects during the expansion phase.
The coalescence procedure is applied at each step of 20 fm/c from t=0 fm/c to follow the evolution of the composite system
up to fragmentation, and deduce properties such as density and excitation energy.
Thus we have access to the corresponding trajectory followed by the composite system 
in the nuclear matter phase diagram and to the properties, in both position and velocity space, of the formed fragments.
The freeze-out time, defined as the time when primary fragments are well formed 
and the nuclear interaction among them is almost negligible,
corresponds to $t\approx 260 fm/c$.
The subsequent propagation in the Coulomb field is treated inside the SIMON code; 20 SIMON decays per SMF event are performed.

\subsection{First stage of the collision : trajectories of the composite system in the phase diagram}\label{Init}

\begin{figure}
	\includegraphics[width=0.99\linewidth]{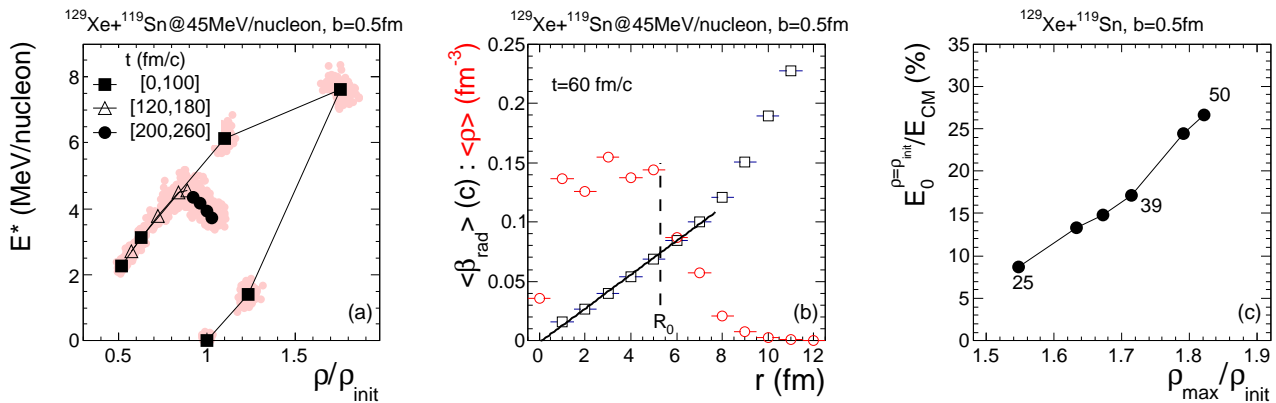}
	\caption{
	(Color online)
	\label{fig:RhoMaxE0}
	\textbf{Left panel}~: mean trajectory in the density ($\rho$) - excitation energy (E*) plane; each black marker stands for a time from t=0 to t=260 fm/c by step of t=20fm/c.
	The markers indicate the mean values while red scatter plots indicate the population for the set of events. \textbf{Middle panel}~: mean evolution of density ($\rho$, red circles) and radial velocity 
	($\beta_{rad}$, black squares) as a function of the distance from the center of the source (r); the continuous line indicates the results of the fit procedure
	of the radial velocity profile (eq.~\ref{eqn:eqnBetaRad}) and the vertical dashed line indicates the value of the root mean square radius of the source (eq.~\ref{eqn:R0}).
	\textbf{Right panel}~: correlation between the radial expansion energy ($E_{0}$, see eq.~\ref{eqn:E0}) and the maximum density ($\rho_{max}$) reached by the source during the compression phase;
	each point stands for each reaction energy from 25 to 50 MeV/nucleon. For normalization of observables see text.
	}
\end{figure}

We address the time evolution of the hot composite sources formed 
in central collisions (b = 0.5 fm).
Such collisions are characterized by a compression phase, due to the initial collisional shock, 
with a subsequent expansion and fragment formation. 
The role of compression and expansion phases
is studied looking at the mean trajectory of the composite source in the excitation energy ($E^{*}$) - 
density ($\rho$) plane. $E^{*}$ represents the thermal excitation energy of the composite source, 
which is evaluated subtracting 
the kinetic energy associated with the Fermi motion (where the Fermi momentum depends on  the local density) 
from the total kinetic energy.
Density is normalized to the value $\rho_{init}$=0.10fm$^{-3}$ corresponding to 
the average density of the colliding nuclei in their ground state.
Such trajectory is plotted on Fig.~\ref{fig:RhoMaxE0}.a, for the reaction at 45 MeV/nucleon.

From these trajectories, we extract the maximum density ($\rho_{max}$) reached by the system, and the maximum 
radial expansion energy ($E_{0}$), corresponding to the time
when the system goes back to normal density, just before the beginning of the expansion phase. 
To extract the radial expansion energy~\cite{Lopez}, we use the radial 
velocity (eq.~\ref{eqn:BetaRad}) profile of the ensemble of test-particles belonging to the composite source 
and apply a fit procedure using eq.~\ref{eqn:eqnBetaRad}.
An example of such a profile is shown on Fig.~\ref{fig:RhoMaxE0}.b. 
After extracting the root-mean-square radius ($R_{0}$, eq.~\ref{eqn:R0}) of the test-particle distribution, 
we then obtain $E_{0}$ (in MeV/nucleon), as written in eq.~\ref{eqn:E0}.

\begin{eqnarray}
	\beta_{rad} = \frac{\vec{\beta}\cdot \vec{r}}{\Vert \vec{r}\Vert}\label{eqn:BetaRad}\\
	\beta_{rad}(r) = a\cdot r + b\label{eqn:eqnBetaRad}\\
	R^{2}_{0} = \sum_{k} A_{k}r^{2}_{k} / \sum_{k} A_{k}\label{eqn:R0}\\
	E_{0} =\frac{1}{2}\;m_{u}\;\beta_{rad}^{2}(r=R_{0})\label{eqn:E0}
\end{eqnarray}

In the equations above $\vec{\beta}$ denotes the local velocity field, $A_k$ is the number of test particles
located at the distance $r_k$ and $m_{u}$ is the unified atomic mass unit.
On Fig.~\ref{fig:RhoMaxE0}.c the extracted values are shown, for all simulated bombarding 
energies (from 25 to 50 MeV/nucleon),
as a function of the maximum density reached by the composite source.
The radial expansion energy is normalized to the available kinetic energy in the center of mass of the 
reaction. The almost linear dependence indicates that to first order the potential energy associated with the compression of the system 
is converted into radial expansion energy in a similar way. However it seems that from $E_{proj}$=39 MeV/nucleon we see a transition with more efficiency
to convert initial compression to subsequent expansion. We also notice that, for the highest energy, 
the radial expansion energy reaches 30\% of the available kinetic energy, which is in good agreement
with experimental estimates deduced from multifragmentation data~\cite{mf}.
The absolute values are also reported in the
table~\ref{tab:RecapE0} (see Section III.C). After this direct evaluation of the collective radial expansion we will now
focus on the production of fragments and its time scale.

\subsection{Expansion phase, nascent fragment partition and recombination effects}

When the hot expanding source reaches low density regions, density fluctuations start to play an important role.
Indeed they are amplified by the unstable mean-field, leading to the possibility to observe multifragmentation.
Multifragmentation is characterized by a simultaneous production of many fragments 
(defined in this work as elements with charge Z$\geq$5).
The excitation energy required for the onset of multifragentation has been derived from comparisons with statistical models~\cite{OnsetLPC,OnsetISIS,Manduci},
and more recently, by measuring the fragment emission time using model-independent Coulomb chronometry~\cite{Fission13}. These studies show that
multifragmentation occurs above $E^{*}$=4 MeV/nucleon, which corresponds to a beam energy of 25 MeV/nucleon for the Xe+Sn central collisions. On the other hand,
in SMF calculations the multifragmentation regime is achieved above $E_{proj}$ = 39 MeV/nucleon.
At lower energies, the fragments produced in the calculations mainly come from secondary decay processes.

\begin{figure}
	\includegraphics[width=0.99\linewidth]{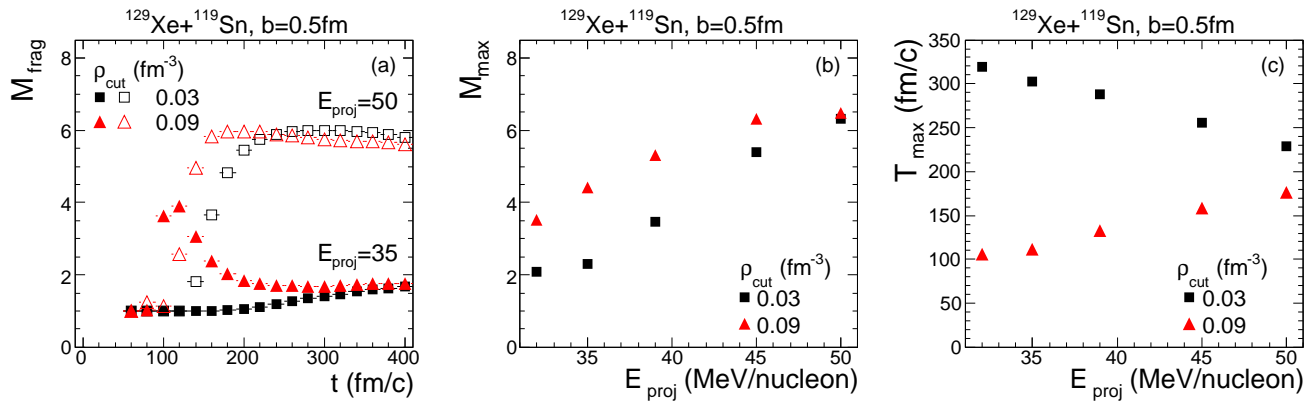}
	\caption{
	(Color online)
	\label{fig:Pattern}
	\textbf{Left panel}~: multiplicity of fragments ($M_{frag}$) time evolution for two density cut-off (black squares and red triangles respectively for $\rho_{cut}$=0.03 and 0.09 fm$^{-3}$)
	and for two beam energies (full and open symbol respectively for $E_{proj}$=35 and 50 MeV/nucleon.); \textbf{middle panel}~: Evolution with beam energies of the maximum value ($M_{max}$) 
	reached by multiplicity of fragment for the two density cut-off; \textbf{right panel}~: evolution with beam energies of time ($T_{max}$) when maximum of fragment multiplicities is reached 
	for the two density cut-off.
	}
\end{figure}

As an illustration, we show on Fig.~\ref{fig:Pattern}.a, the mean evolution with time of the multiplicity 
of primary fragments ($M_{frag}$) identified through the coalescence procedure,
for two beam energies, 35 MeV/nucleon (full symbols) and 50 MeV/nucleon (open symbols), respectively below and above this threshold. 
For each beam energy, the two curves stand for two values of the density cut ($\rho_{cut}$) used in the coalescence procedure : the usual one (0.03$fm^{-3}$, black squares) and a greater one (0.09$fm^{-3}$, red triangles), 
aiming at an easier recognition of fragments in the case of compact configurations.

Let us consider the usual density cut (0.03$fm^{-3}$). At 35 MeV/nucleon, the mean fragment multiplicity increases slowly up to $M_{frag}\sim 1.7$. At 50 MeV/nucleon, fragments
are produced quickly and the multiplicity reachs its maximum ($M_{frag}\sim 6$) around t=240fm/c. To go more into details of the fragment production during the dynamical stage, we now investigate
the evolution of $M_{frag}$ with $\rho_{cut}$=0.09$fm^{-3}$. At 50 MeV/nucleon, a similar evolution is observed but faster by about 60-80 fm/c.
The main difference concerns $E_{proj}$=35 MeV/nucleon : high values of $M_{frag}$ compatible with multifragmentation are observed around 
t=120 fm/c (i.e. on short time scales). Then the multiplicity decreases and reaches the values corresponding to $\rho_{cut}$=0.03$fm^{-3}$ at larger times. 
This indicates that, during the expansion phase, some pre-fragments are formed, but they do not survive and recombine, leading to the observation of one or two big fragments
in the exit channel.

This transient state consisting of nascent partitions which then recombine is observed from $E_{proj}$=32 MeV/nucleon. To extract its properties, we search event by event 
for the time ($T_{max}$) where the number of fragments reaches its maximum value 
or saturates ($M_{max}$) and then compute the mean values~\footnote{It has to be noted that
to compute maximum multiplicity and associated time only events having, at a given time, two or more fragments, are taken into account.}.
On figures \ref{fig:Pattern}.b and \ref{fig:Pattern}.c,
the evolution of these two quantities are shown for the two $\rho_{cut}$ values for $E_{proj}$ between 32 and 50 MeV/nucleon.
We see that using $\rho_{cut}$=0.09$fm^{-3}$, significant values of the multiplicity are achieved (3.5-6) even at low energy, corresponding to nascent fragment partitions.
For the standard $\rho_{cut}$=0.03$fm^{-3}$, multifragmentation events are dominant above 39 MeV/nucleon. 
Indeed, increasing the beam energies, we observe closer values for the two density cuts. 
On Fig.~\ref{fig:Pattern}.c, the time ($T_{max}$) associated with the maximum multiplicity is plotted. 
It exhibits an opposite trend for the two cut-offs and values converge with increasing beam energies.
For $\rho_{cut}$=0.09 $fm^{-3}$ and lower energies, very short times are observed, indicating that the recombination 
of nascent partitions is rather fast. On the other hand, at higher energies time scales are longer,
but coherent with the picture of a slower continuous evolution towards final fragment partitions 
in the exit channel.

We conclude that density fluctuations in the low density composite system are sufficient to lead to pre-fragment formation
at early times. From this common stage, two scenarios are observed : nascent partitions recombine into a composite source 
or survive leading to multifragmentation. The amount of collective radial expansion appears to be the main discriminator.
The shift of the multifragmentation threshold towards higher beam energies, as compared to experimental results, 
indicates an underestimation of the radial expansion. As it will be shown in the next Section, 
fragment velocities are underestimated also when multifragmentation is observed.
These observations are linked to
known limitations of semi-classical models, with respect to a full quantum treatment :
indeed semi-classical models are characterized by a reduced capability of the excited nuclear systems to keep
particles inside the potential well, due to the lack of quantum reflection effects~\cite{Lacroix}.
Instead of expanding under the effect of the thermal pressure, the hot system generally emits a large number of particles,
reducing its excitation energy and temperature.
To quantify the damping of collective flow, we go now into a realistic comparison between SMF+SIMON outputs
and INDRA multifragmentation data.

\section{Comparison to experimental data}

\subsection{Simulated cross section and freeze-out time.}

We now present a comparison between SMF calculation outputs
and INDRA data obtained from selected central collisions. 
We will seak, in the experiment and in the simulations, for the 
events which correspond to the disassembly of a unique
composite source (denoted as ``fusion'' events in the following).
To do so, we consider SMF simulations mainly in the impact parameter ($b$) range
between 0.5 and 4 fm. In the following we will show that for beam energies above 39 MeV/nucleon, this range is sufficient to select all fusion events while
for lower energies, fusion events contribute also at greater impact parameter.
Starting from a flat $b$ distribution and taken the total reaction cross section given by the Kox formula~\cite{Kox}, we apply a renormalisation
to mimic the typical triangular distribution of impact parameter.
For the considered impact parameter range, it corresponds to around 10\% of the reaction cross section~\footnote{The reaction cross section values are in the range [5.7;6.0] barn, with an associated maximum impact parameter 
in the range [13.4,13.8] fm.}.
As stated above, we consider that freeze-out is achieved at time t=260 fm/c. A this time, for fusion events at beam energies above 39 MeV/nucleon,
fragment production and relaxation in momentum space is almost achieved in SMF.
Excitation energy ($E^{*}$) and density ($\rho$) of primary fragments are independent of their charge with mean values 
between 3.4 and 3.8 MeV/nucleon for $E^{*}$ and 1.0 and 1.1 for
$\rho/\rho_{init}$. Angular momentum values are between 0 and 30 $\hbar$.
In SIMON code, primary hot fragments are supposed spherical and at normal density. These asumptions are thus reasonable for fusion events 
at beam energies above 39 MeV/nucleon. 
For the lowest energies, the time evolution indicates that SMF events which do not break into pieces are not completely 
relaxed in ${\bf r}$ space and the passage between SMF and SIMON could lose some coherency. 
Anyhow for the present work, the main goal is to compare with experimental data the cases when
complete multifragmentation is achieved and we take t=260fm/c as freeze-out time for all energies and all impact parameters.
For each SMF event, we perform 20 SIMON de-excitations and obtain a comparable set of data to study central collisions.
The first step is to test the assumptions made in the data to select the so-called central collisions.

\subsection{Selection of central collisions in SMF+SIMON}

To study fragmentation properties of hot primary sources in experimental data,
one has to focus on the most central collisions (low impact parameter range) and try to extract them from the whole set of collected reactions.
Before that, a preliminary selection has to be made on the total detected charge ($Z_{tot}$, eq.~\ref{eqn:Ztot}) in the experimental events. 
We keep only the so-called complete 
events, with $Z_{tot}$ greater or equal than 90\% of the total charge of the system (104). Fragment properties of the selected set of events are 
compared direclty to the results  of the simulations, without filtering.
We will focus in the following on the selection based on the orientation of events in momentum space which has been widely used
by the INDRA collaboration~\cite{Marie,Hudan,Rivet}.
We calculate event by event the kinetic energy tensor (eq.~\ref{eqn:Tenseur}) of fragments. From its eigen values, an ellipsoid is defined which describes the matter distribution.
In this way the flow angle ($\theta_{flow}$, eq.~\ref{eqn:CosTheta}), is defined as the angle between the main axis of the 
ellipsoid $\vec{e_{3}}$ and the beam direction ${\vec{k}}$. 

\begin{eqnarray}
	Z^{tot} = \sum^{Mtot}_{i=1} Z_{(i)}\label{eqn:Ztot}\\
	Q_{ij} = \sum_{\nu=1}^{Mfrag} \frac{p_{\nu}^{(i)}p_{\nu}^{(j)}}{m_{\nu}(\gamma_{\nu}+1)}\label{eqn:Tenseur}\\
	\theta_{flow} = \arccos(\mathbf{\vec{e_{3}}} \cdot \mathbf{\vec{k}})\;\;\in\;\;[0;90^\circ]\label{eqn:CosTheta}
\end{eqnarray}
In the equation, $m_\nu$ and $p_\nu$ denotes the fragment mass and momentum, and $\gamma$, the Lorentz factor.
The main justification of the use of this observable is the following : an equilibrated system which undergoes multifragmentation 
will produce an isotropic distribution of fragments leading to a flat distribution for the $cos\;\theta_{flow}$ observable while
a system which keeps a memory of the entrance channel favours orientation along the beam axis and will produce a distribution peaked at $cos\;\theta_{flow}$=1.

\begin{figure}
	\includegraphics[width=0.99\linewidth]{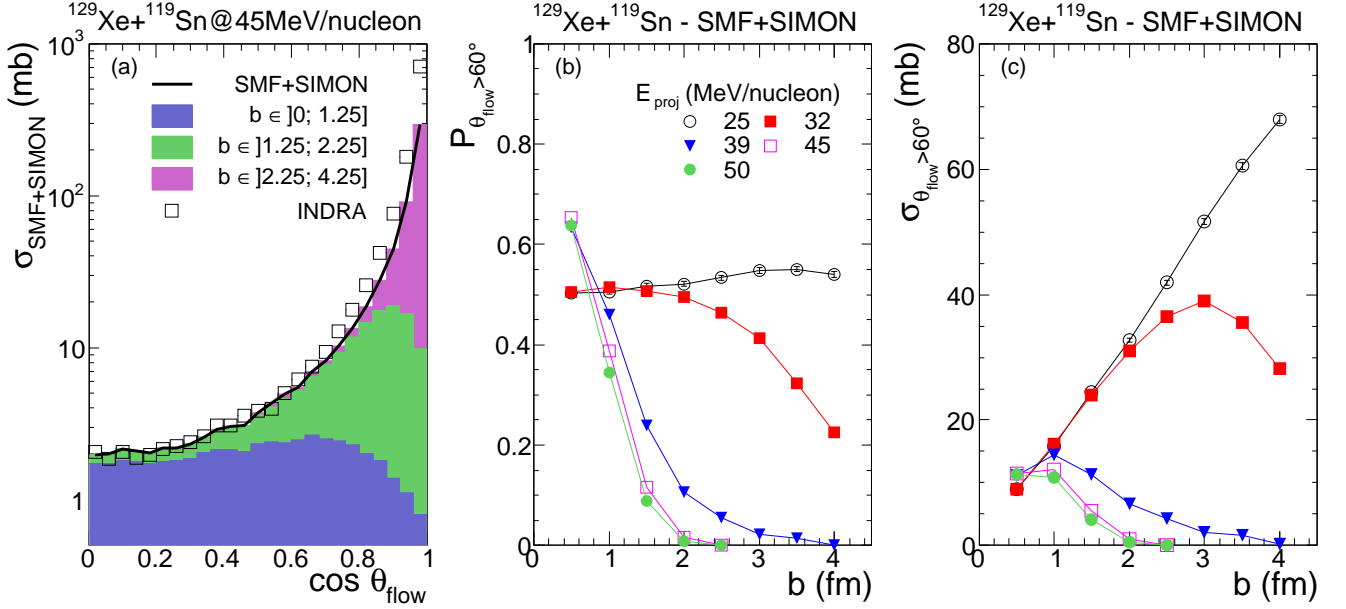}
	\caption{(Color online)
	\label{fig:ThetaFlow}
	\textbf{Left panel}~: distributions in mb of the flow angle cosinus ($cos\;\theta_{flow}$) distribution : black line distribution is for SMF+SIMON events and the three colors stand for the relative contribution 
	of 3 impact parameter (b) ranges. Black squares distribution is for INDRA data after normalization to the simulated cross section (see text for details); \textbf{middle panel}~: probability for each impact 
	parameter (b) that simulation events fulfil the condition $\theta_{flow}>60^{\circ}$; \textbf{right panel}~: cross section contribution of each impact parameter to the selected sample of events.
	}
\end{figure}

On Fig.~\ref{fig:ThetaFlow}.a flow angle distributions are shown for SMF+SIMON (black line) and INDRA events (black squares) for
$E_{proj}$=45 MeV/nucleon.
Data are normalized to the distribution of simulated events, using the ratio of the cross section of a posteriori selected events (see table~\ref{tab:CrossSection}). The good agreement of the shape indicates
a good reproduction of the topology of events by SMF. The distribution is made from two components : the most central collisions (blue area) produce a rather flat
distribution, while a more forward-peaked distribution is due to more peripheral collisions (green and purple areas).
In the experimental data, binary collisions are rejected by the requirement on the total 
detected charge. This is  the reason why considering impact parameters below 4 fm in the simulations is sufficient to reproduce the experimental flow angle distribution.
In the following, we apply the same criterion to the simulation outputs as for experimental data : $\theta_{flow}>60^{\circ}$.\\

On figures \ref{fig:ThetaFlow}.b and \ref{fig:ThetaFlow}.c, effects of the selection on the simulated events statistics in terms of probability and cross section are shown. Looking first at 
evolution of probabilities with impact parameter, two regimes appear located below and above 39 MeV/nucleon. For higher energies, the probability that events are selected becomes negligible 
above b=1.5fm indicating a rather sharp transition between fusion and binary collisions. Moreover it has to be noticed, that for b=0.5fm, events with perpendicular orientation with respect to the beam axis ($cos\;\theta_{flow}$=0) 
are favoured, indicating a full stopping.
Concerning lower energies, fusion events are also produced in mid-central reactions (b$\in[3;6]$ fm) which explain the important part of selected events.
Indeed mean-field based approaches tend to overestimate mean-field dissipation and
orbiting effects. Instead of reseparating after a short interaction time, projectile and target stay stuck and produce high spin composite systems. 
For $E_{proj}$=39, 45, 50 MeV/nucleon, the chosen impact parameter
range b$\in[0.5;4.]$fm is sufficient to collect all events which fulfil the condition $\theta_{flow}>60^{\circ}$ while for $E_{proj}$=25 and 32 MeV/nucleon, a wider
range is necessary~\footnote{For $E_{proj}$=25 MeV/nucleon, the maximum of the selected cross section distribution is located at b=5 fm.}.

\begin{table}
	\begin{tabular}{c||c|c||c|c|c}
	\hline
	 & \multicolumn{2}{|c||}{$\sigma_{\theta_{flow}>60^{\circ}}$ (mb)} & \multicolumn{3}{|c}{$P_{ M_{frag\geq3} }$} \\
	$E_{proj}$ & SMF+SIMON & INDRA & SMF & SMF+SIMON & INDRA\\
	\hline
	25 & {\it39.3}	& 18 	& 0.0 & 0.64 & 0.94\\
	32 & {\it30.4}	& 4.3	& 0.0 & 0.58 & 0.98\\
	39 & {\it3.3}	& 2.3	& 0.45 & 0.73 & 0.98\\
	45 & {\it 1.9} (30.3)	& 1.9 & 0.98 & 0.98 & 0.99\\
	50 & {\it1.7} 	& 1.8	& 1.00 & 1.00 & 0.98\\
	\end{tabular}
	\caption{
	\label{tab:CrossSection}
	\textbf{Second and third columns}~: Cross sections in mb of selected SMF+SIMON and experimental INDRA events. Values for SMF+SIMON are scaled to the experimental value at $E_{proj}$=45 MeV/nucleon (see text for
	details); \textbf{three last columns}~: probabilities that multiplicity of fragments are greater or equal to 3 for SMF primary events, SMF+SIMON final events and
	experimental INDRA events.
	}
	
\end{table} 

In first part of table~\ref{tab:CrossSection}, we report the total cross section of the selected events, as observed 
in the data and in the simulations. Since only almost complete events are retained in the experimental data analysis, the associated cross section depends on the detection efficiency.
Cross sections in the simulations, have to be normalised to get comparable values. We consider the beam energy $E_{proj}$=45 MeV/nucleon, where 
the simulations (SMF+SIMON) 
provide a good description of the data : 
from the corresponding cross section values, in  the data (1.9 mb) and in the simulations~(30.3 mb), we define a normalisation factor as 
the ratio between the two. 
This factor is used in figures \ref{fig:ThetaFlow} and \ref{fig:Comparaison}. After this renormalization, cross section values, for the highest energies, 
are close to 2 mb.
For lower energies,
the overestimation of simulated cross sections is coherent with the contribution coming from mid-central reactions mentioned before~\footnote{The overestimation is less important for 25 MeV/nucleon than 32 MeV/nucleon, 
because for 25 MeV/nucleon, INDRA detector is more efficient to detect the so-called complete events.}.\\
The first step to evidence multifragmentation processes in the simulations is the presence of events with 
$M_{frag}\geq3$, at the primary level. We report in the second part of table~\ref{tab:CrossSection}, the probability that events 
fill this criterion ($P_{ M_{frag\geq3} }$). The fourth and fifth columns stand for SMF primary events and SMF+SIMON final events. The whole fragment production for $E_{proj}$=25 and 32 MeV/nucleon comes from sequential splitting
calculated by the SIMON code. The greater values of $P_{ M_{frag\geq3} }$ for 25 MeV/nucleon, with respect to 32 MeV/nucleon, is due to the exploration of high spin regions by the composite system in the mid-central reactions.
Indeed such exotic events are less important for $E_{proj}$=32 MeV/nucleon. For $E_{proj}$=39 MeV/nucleon, half of events are multifragmentation events while for greater energies full multifragmentation is achieved.
The last column reports the same information for INDRA data : the probability is close to 1 on the whole considered beam energy range.

\subsection{Fragment charge and velocity distributions for $E_{proj}$=39, 45, and 50 MeV/nucleon.}

In the following we will compare fragment properties, as given by the SMF + SIMON simulations, to the 
INDRA data. We will focus on the comparison for the beam energies 39, 45 and 50 MeV/nucleon, 
where SMF calculations lead to multifragmentation events.
\begin{figure}[!hhhh]
	\includegraphics[width=0.99\linewidth]{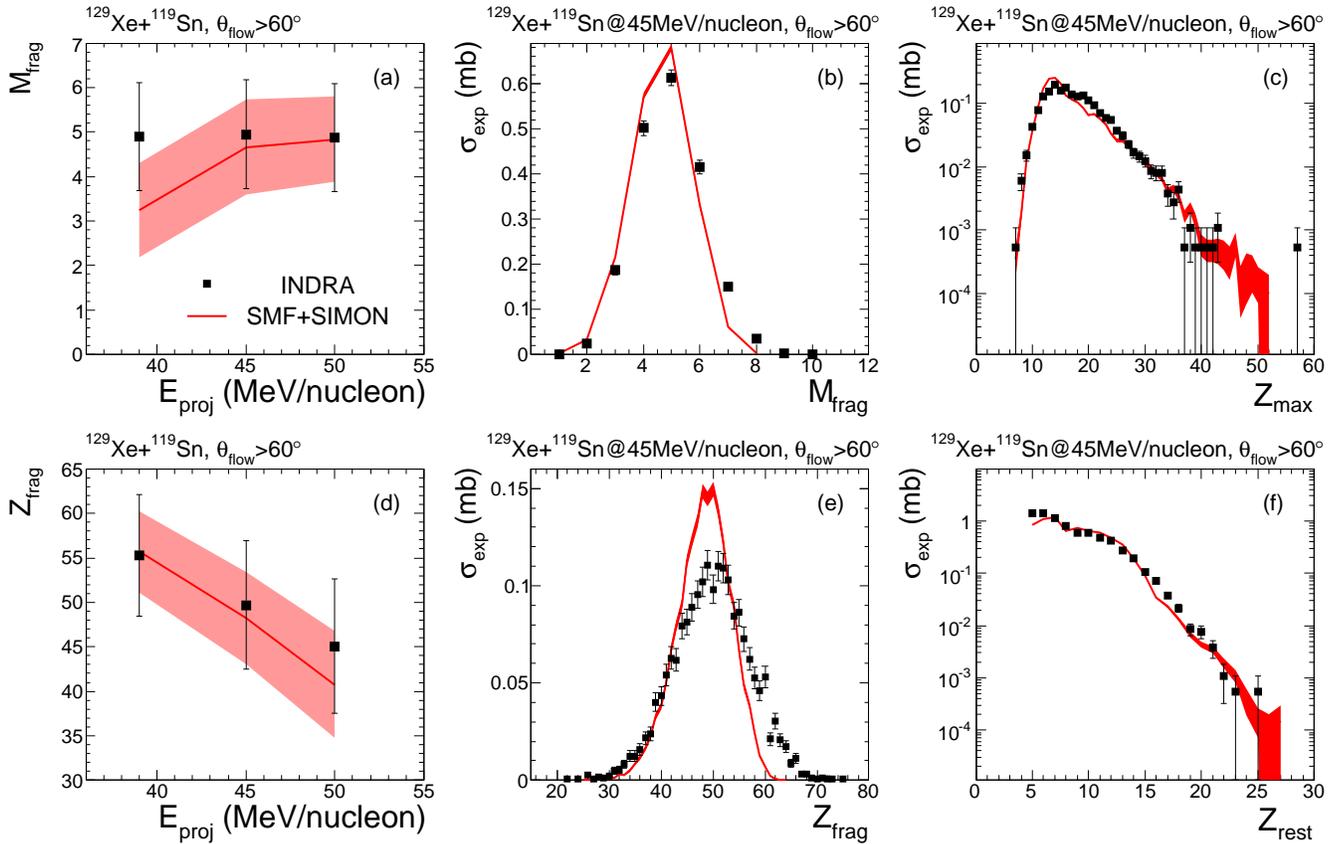}
	\caption{
	(Color online)
	\label{fig:Comparaison}
	Comparison between INDRA (black full squares) and SMF+SIMON (full red line) data. The same $\theta_{flow}$ cut is applied to select central collisions. \textbf{Left panels}~: Evolution with beam energy ($E_{proj}$) 
	of the mean values of fragment multiplicity ($M_{frag}$, top) and charge bound in fragments ($Z_{frag}$, bottom); error bars indicate the standard deviation of distributions. \textbf{Middle panels}~: Distribution of the fragment multiplicity ($M_{frag}$, top) and charge bound in fragments 
	($Z_{frag}$, bottom). \textbf{Right panels}~: Charge distribution of the biggest fragment ($Z_{max}$, top) and of the other fragments ($Z_{rest}$, bottom). 
	For the experimental distributions, the Y-axis scale is in mb, and calculations are scaled on it.
	}
\end{figure}

Figures \ref{fig:Comparaison}.a and \ref{fig:Comparaison}.d show the evolution of final fragment multiplicity ($M_{frag}$)
and the total charge bound in fragments ($Z_{frag}$) with beam energy. The error bars indicate the standard deviation of the distributions.
At $E_{proj}$=39 MeV/nucleon, we see that multiplicity is still underestimated in SMF but becomes fully compatible with data for higher energies.
The average $Z_{frag}$ values are in good agreement with the data
at all energies. However, it should be noticed that more stringent constraints on the completeness of the selected events could lead  to larger experimental values.
 
For both observables, the width of the distribution is underestimated in SMF and the general trend is that the high value tail of the distributions is less well 
reproduced (see figures \ref{fig:Comparaison}.b and \ref{fig:Comparaison}.e, where these distributions are shown for $E_{proj}$=45 MeV/nucleon). 
This indicates that the most explosive events are less present in the SMF calculations, pointing to a too intense primary nucleon emission during 
the fragmentation process.
On figures \ref{fig:Comparaison}.c and \ref{fig:Comparaison}.f are plotted the charge distribution of the biggest fragment ($Z_{max}$) and of all the other fragments ($Z_{rest}$), at 45 MeV/nucleon, as obtained in the simulations and in the data. 
The comparison is rather good, showing that the sharing of charge among fragments is well managed by the SMF approach. This feature is all the more important, as
the good reproduction of the whole distribution of the charge of the biggest fragment is mandatory in the study of signature of phase transition
or critical phenomena done in experimental analyses\cite{Bimod,Gruyer}.\\

\begin{figure}
	
	\includegraphics[width=0.99\linewidth]{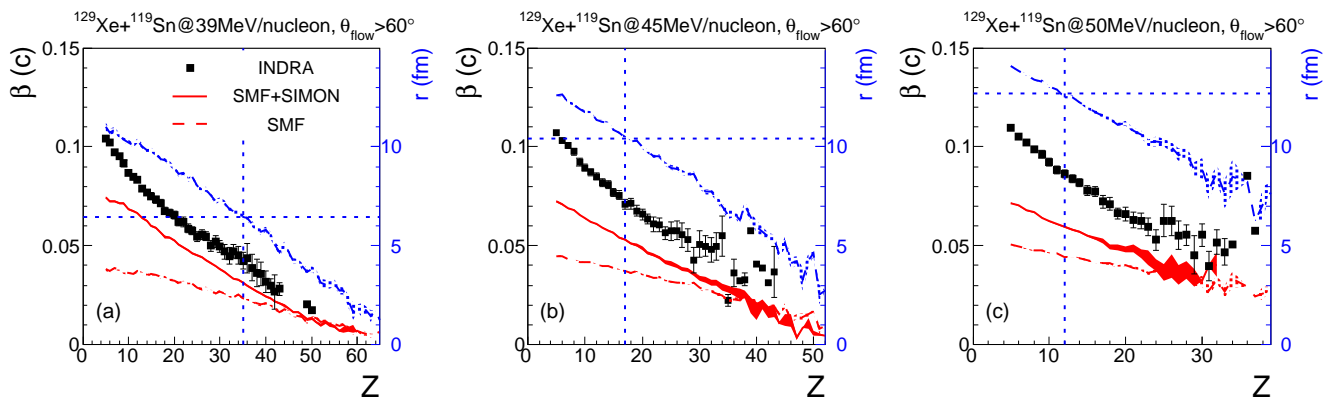}
	\caption{
	(Color online)
	\label{fig:BetaZ}
	\textbf{From left to right}~: energies 39, 45 and 50 MeV/nucleon, the same $\theta_{flow}$ cut is applied to select central collisions. Evolution of the mean values of velocity ($\beta$ in c unit) 
	of fragments with respect to their charge (Z) : black full squares are for INDRA data, dashed red line for SMF data at t=260fm/c and full red line for SMF+SIMON data; in addition the 
	blue line shows the evolution of the mean values of the distance of fragments (r in fm, right axis) with respect to their charge (Z). The horizontal dashed blue line indicates the value 
	of the root mean square radius of fragment distribution in r space ($R_{0}$ in fm) at t=260 fm/c and the vertical line indicates the value of the charge of the fragments
	located, on average, at this position.
	}
\end{figure}

In the following we discuss fragment kinematic properties, focusing on the radial flow, which, as it was shown in~\cite{Moisan}, is underestimated in
SMF calculations. In order to undertake a quantitative analysis, we present in figure~\ref{fig:BetaZ}, the fragment mean velocity as a function of the fragment charge. 
Three profiles, calculated before (t=260 fm/c in SMF) and after the Coulomb propagation (SMF+SIMON) and finally the experimental one (INDRA), are displayed.
We also represent the average radial distance of fragments, evaluated  at t = 260 fm/c. The hierarchy observed for 
the radial distance and the velocity is the same : the largest charge is associated with the smallest distance and velocity. The velocity and
distance trends have a similar slope after the Coulomb propagation, whereas the slope of the SMF velocities is lower,  
indicating the impact of Coulomb effects on the
final (SMF+SIMON) distribution. From the radial distance and velocity SMF profiles, 
the radial expansion energy ($E^{SMF}_{o}$, eq.~\ref{eqn:E0}) 
can be extracted, using the same method described in Section 2.3.
Values are reported in tab.~\ref{tab:RecapE0} (second column). The comparison with values measured at the early stage of the expansion 
(first column) indicates a decrease by a factor between 4 and 8, depending on the beam energy. 
The values of the root-mean-square radius ($R_{0}$, eq.~\ref{eqn:R0}), are also reported.  
To estimate the missing radial energy ($\delta{E_{0}}$) in the calculations, we adopt the following procedure.
We consider the average charge corresponding to the distance $R_{0}$ (the horizontal and vertical dashed lines in
fig.~\ref{fig:BetaZ} indicate these values) and we evaluate the Coulomb contribution to the radial energy as
$E^{(clb)}=E^{(SMF+SIMON)}_{kin} - E^{(SMF)}_{0}$, where $E^{(SMF+SIMON)}_{kin}$ is the final kinetic energy, per nucleon, for 
the charge considered. As the trend of velocity profiles for SMF+SIMON and INDRA data is similar, charge partitions are well reproduced and fragment spatial distribution
at freeze-out  is well described by SMF~\cite{Tabacaru}, thus the final Coulomb repulsion should contribute in the same way to the fragment velocity spectra.
Then, knowing  $E^{(clb)}$, from the experimental kinetic energies one can deduce the corresponding kinetic energy at freeze-out and
estimate  the missing radial energy ($\delta{E_{0}}$) in SMF. The values obtained are also reported  
in tab.~\ref{tab:RecapE0}. Estimates of radial energies performed by Piantelli and co-workers~\cite{Piantelli} are added in the last column of the table. 
It is worth noting that the simulation developed in~\cite{Piantelli} is based on the same experimental data
and allow one to constrain the freeze-out configuration, treating the radial expansion  as a parameter. 
We observe that the values obtained  in~\cite{Piantelli} are in good agreement with the $\delta{E_{0}}$ derived from  our analysis. 

Putting together all the results shown in Sections 2 and 3, we clearly observe in the SMF calculations a transition between sequential de-excitation and full multifragmentation around 39 MeV/nucleon. 
At lower energies, the radial expansion energy is not sufficient to bring the observed nascent partitions
to a final multifragmentation of the system. At the same energies, fusion and orbiting effects are overestimated at larger impact parameters.
Two features can explain this behavior. The additional expansion driven by the thermal pressure is not well treated. 
Secondly (but related to this), the expansion phase of the system, where fluctuations should develop, could
be too dissipative in terms of nucleon emission.
For instance, for the reaction at 50 MeV/nucleon we observe that our calculations overestimate, by about a factor 2, the proton yield, due to more abundant pre-equilibrium nucleon emission, whereas the yield of charges $Z = 2-4$ is
underestimated by a factor 2. This problem can only be partially cured by an improved description of many-body
correlations and fluctuation dynamics in semi-classical approaches \cite{BLOB}.
Anyhow, we believe that the analysis considered here, which couples a detailed study of ingredients and limitations of the model to a realistic
comparison of observables, should improve our comprehension of the role of the radial expansion energy along the multifragmentation process.

\begin{table}
	
	\begin{tabular}{c||c||c|c||c|c}
	\hline
	$E_{proj}$ & $E_{0}^{ \rho=\rho_{init}}$ (MeV/nucleon)& $E^{ SMF}_{0}$ (MeV/nucleon)& $R^{ SMF}_{0}$ (fm) & $\delta E_{0}$ (MeV/nucleon)&~\cite{Piantelli} (MeV/nucleon)\\
	\hline
	39 & 1.65 & 0.26 & 6.4 & 0.39 & $0.9^{+0.45}_{-0.45}$\\
	45 & 2.73 & 0.65 & 10.4 & 1.04 & $1.1^{+0.55}_{-0.55}$\\
	50 & 3.30 & 0.92 & 12.7 & 1.82 & $1.7^{+0.20}_{-0.55}$
	\end{tabular}
	\caption{
	\label{tab:RecapE0}
	\textbf{From left to right}~: expansion energy at the time when the system goes back to normal density, after the compression phase (see section~\ref{Init} for details); SMF calculation at t=260fm/c : expansion energy with associated root mean square radius of fragment distribution 
	in r space; estimation of missing expansion energy using experimental fragment velocities as reference; \textbf{the last column} displays the expansion energy values added as a free parameter in
	simulation done in~\cite{Piantelli}.
	}

\end{table} 

\section{Stopping power analysis: comparison between simulations and data}

\begin{figure}
	
	\includegraphics[width=0.99\linewidth]{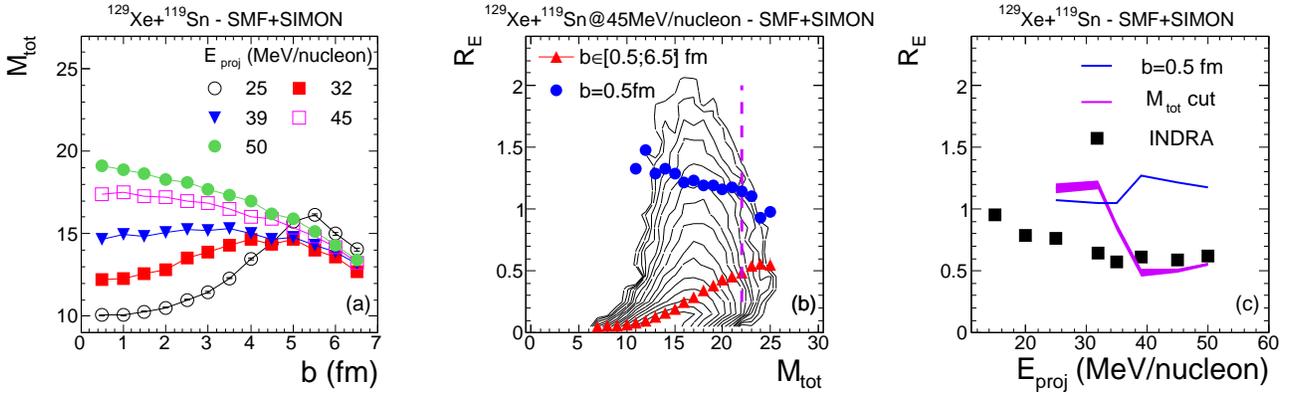}
	\caption{
	(Color online)
	\label{fig:Stopping}
	\textbf{Left panel}~: evolution of total multiplicity ($M_{tot}$) with the impact parameter; \textbf{middle panel}~: for $E_{proj}=45 MeV/nucleon$, correlation between the isotropic ratio ($R_{E}$) and $M_{tot}$ in black contours (logarithmic scale),
	the full red triangles are the deduced profile. The profile for b=0.5 fm is plotted in blue circles. The vertical violet dashed line indicate where the $M_{tot}$ cut is performed (see text for details); \textbf{right panel}~: evolution 
	with beam energies ($E_{proj}$) of mean values of $R_{E}$ for b=0.5fm (blue line) and for events selected using the $M_{tot}$ cut (violet line). INDRA data, taken from~\cite{Lehaut}, are plotted in black squares.
	}
	
\end{figure}

The stopping power in nuclear collisions measures the efficiency of conversion of the initial beam energy into transverse
directions and it is a useful tool to investigate the interplay between mean-field (one-body) properties and two-body correlations such as nucleon-nucleon collisions.
This observable has been experimentaly investigated at relativistic~\cite{Reisdorf} and intermediate~\cite{Lehaut} beam energies.
At relativistic energies~\cite{Reisdorf}, maximum stopping is reached between 0.2 and 0.8 GeV/A for heavy systems (Au+Au), with a shift
to higher energies for smaller systems. In coincidence, a maximum side-flow is also measured, making coherent the picture of the initial beam 
energy converted into transverse energy.\\
We will now discuss the stopping analysis on the Xe+Sn system at intermediate energies. 
The authors of Ref.\cite{Lehaut} aimed at measuring the stopping power in
nuclear systems formed in heavy ion symmetric collisions. They adopt the isotropy ratio ($R_{E}$, eq.~\ref{eqn:Riso}) as related observable
and use the total 
charged particle 
multiplicity ($M_{tot}$) to sort the events according to their degree of dissipation~\footnote{In this work and related ones, the 
isotropic ratio is also computed with the linear momentum : $R_{p}= 2/\pi \sum_{i=1}^{Mtot} p^{(i)}_{\perp}/\sum_{i=1}^{Mtot} p^{(i)}_{//}$. To keep it simple, we only use in the present analysis
the $R_{E}$ ratio and discuss results and comparisons assuming that these two ratios give the same quantitative results.}:
\begin{eqnarray}
	R_{E} = \frac{1}{2} \frac{\sum_{i=1}^{Mtot} E^{(i)}_{\perp}}{\sum_{i=1}^{Mtot} E^{(i)}_{//}}\;\;\in\;\;[0;\infty[
	\label{eqn:Riso}
\end{eqnarray}
where $ E_{//}$ and $E_{\perp}$ denote, respectively,  the parallel and transverse energy of the detected particles.
For each beam energy, the events corresponding to the largest multiplicity
($M_{tot}$) region, where $<R_{E}>$ is rather constant, are considered. Then the associated average $R_{E}$ value is extracted. 
The evolution of $R_E$, as a function of the beam energy, is reported on Fig.~\ref{fig:Stopping}.c 
(black full squares). Starting at $E_{proj}$=12 MeV/nucleon, a $R_{E}$ value close to 1 is observed, indicating a complete stopping,
whereas, increasing the beam energy, the ratio decreases and reaches a minimum, around 0.55, at 40 MeV/nucleon. A small increase to 0.6, 
at 100 MeV/nucleon, is finally observed. 
From these results, it was argued that, in the Fermi energy domain, stopping is not achieved 
because mean-field dissipation is not sufficient
and that, starting from 
40 MeV/nucleon, the effect of in-medium nucleon-nucleon collisions becomes predominant, increasing the 
degree of stopping.\\
Recently, results from IQMD calculations, performed for the same reactions, have been reported in~\cite{GQ_Zhang}.
The authors apply the same selection, as considered in the data analysis, to the simulated events,
and show that this leads to impact parameter mixing. Moreover, in Ref.\cite{GQ_Zhang} it was stressed that fragment configurations, rather than the nucleon phase space, have to be used 
in order to compare with the experimental results. Such link between clusterization and stopping evaluation is also reported in~\cite{Y_Zhang}.
Finally, a qualitative agreement with the general trend of the data is observed, however the minimum of $R_{E}$ appears at higher energies (with a shift of 20 MeV/nucleon) and the rise at beam energies of 100-150 MeV/nucleon is more pronounced.\\

In the following we will present the results obtained within the SMF + SIMON approach. 
From the flow angle distribution studied in the previous Section (Fig.~\ref{fig:ThetaFlow}), we have already observed 
favored prolate orientation for events at small impact parameters, indicating the presence of full stopping and 
a loss of the entrance channel memory. To complete this observation, we perform an analysis similar to the one in~\cite{Lehaut}, 
on the impact parameter range [0.5; 6.5] fm.

First, on Fig.~\ref{fig:Stopping}.a, we show the evolution of $M_{tot}$ (which excludes pre-equilibrium particles) 
with the impact parameter (b). A different behavior is observed for energies below and above 39 MeV/nucleon with first an increasing and then 
a decreasing trend of $M_{tot}$ with respect 
to b. At small beam energies, the largest $M_{tot}$ is observed at large impact parameters ($b=5-6~ fm$), corresponding to high spin sequential splitting,
whereas the opposite holds at high beam energies.
We notice that at the ``transition energy'' (39 MeV/nucleon) we also 
observe a transition, at small impact parameters, from sequential decay of heavy residue to multifragmentation events. 
Fig.~\ref{fig:Stopping}.b shows the correlation between $R_{E}$ and $M_{tot}$, 
as proposed~in~\cite{Lehaut}, at $E_{proj}$=45 MeV/nucleon. We observe the same trend, as seen in the data, 
for the global population of the diagram (black contour plot) and for the mean (red full triangles) : 
saturation of $R_{E}$ for the highest $M_{tot}$ values and large fluctuations for multiplicities around
half of the maximum reached value. 
However,  this picture is valid for beam energies above 39 MeV/nucleon, whereas, at lower energies, 
the trend of $R_{E}$ as a function of $M_{tot}$ is rather flat, in the considered impact parameter range. 
Indeed low multiplicity corresponds to small impact parameters
(see Fig.~\ref{fig:Stopping}.a) and at larger impact parameters (large $M_{tot}$ in this case)
the dynamics is still rather dissipative, thus $R_{E}$ is around 1 in all cases. 

\begin{figure}
	
	\includegraphics[width=0.99\linewidth]{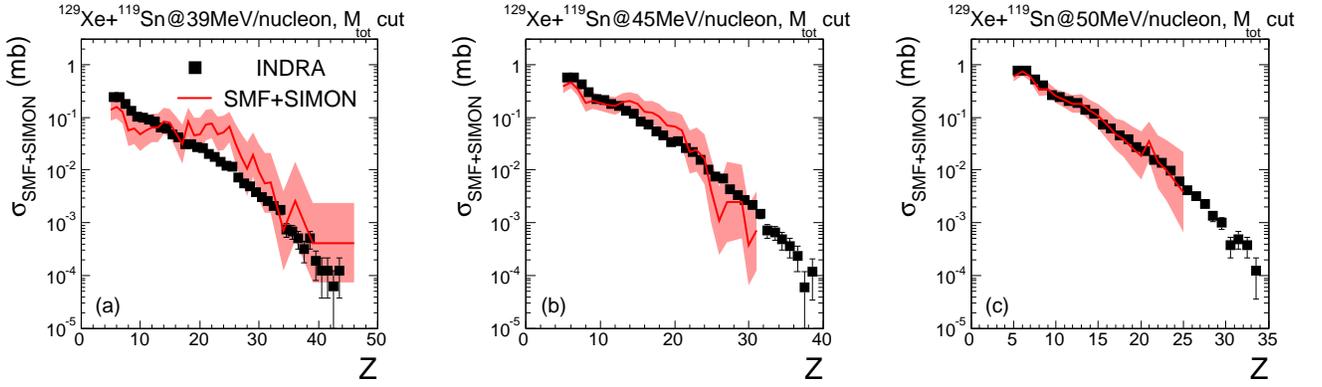}
	\caption{
	(Color online)
	\label{fig:DistriZMtotCut}
	\textbf{From left to right}~: Energies 39, 45 and 50 MeV/nucleon, the $M_{tot}$ cut is applied to select the most dissipative collisions. 
Charge (Z) distribution: black full squares are for INDRA data and full red line for SMF+SIMON calculations.
	For the simulated distributions, the Y-axis scale is in mb, and experimental data are scaled on it.
	}
	
\end{figure}

We also plot on Fig.~\ref{fig:Stopping}.b the same correlation  just considering b=0.5 fm (blue points). We observe that a wide $M_{tot}$ region is populated
and that $R_{E}$ values are greater than 1, indicating that full stopping is achieved for the most central collisions~\footnote{The computation 
of $R_{E}$ for b=0.5fm case is done on all events while for the other cases it
has been computed event by event. This allows to avoid spurious fluctuations due to the ratio between small numbers, especially at 
lower beam energies where, in the majority of events, there is a big residue located near the center of mass.}.

On Fig.~\ref{fig:Stopping}.c, we report the $R_{E}$ mean values, corresponding to b=0.5 fm, for all beam energies (blue lines). 
An almost flat behaviour is observed. 
To perfom a similar $M_{tot}$ cut in the SMF events, 
applicable to the whole beam energy range, we consider the value of $M_{tot}$ above which
the cross section of the selected events is equal to the value associated with b=0.5 fm (which corresponds 
to 3\% of the simulated cross section).
On Fig.~\ref{fig:Stopping}.b, the violet vertical line indicates the $M_{tot}$ cut obtained at $E_{proj}$= 45 MeV/nucleon . The violet line on Fig.~\ref{fig:Stopping}.c 
represents the evolution of $R_{E}$ with beam energies, according to this selection. 
Results from~\cite{Lehaut} are reported as black full squares.
At the lowest energies, values are still close to one indicating important dissipative and orbiting effects at large impact parameters.
Then, increasing the beam energy up to 39 MeV/nucleon, $R_{E}$ decreases, but this is due to the impact parameter mixing and to the
contribution of intermediate impact parameters.
In fact, as one can see on Fig.~\ref{fig:Stopping}.a, at 39 MeV/nucleon the total multiplicity exhibits a rather flat
behavior with b. Clearly, for mid-central collisions,
the reaction dynamics becomes more transparent, leading to a reduction of $R_{E}$.
The impact parameter mixing is less pronounced at beam energies larger than 39 MeV/nucleon, where the total multiplicity
decreases significantly with b. This leads to the small increase observed for $R_{E}$ on Fig.~\ref{fig:Stopping}.c.  
In the low beam energy region, the results overestimate the data due to the overestimation of mean-field dissipation.
In the high beam energy region, where complete multifragmentation is achieved, SMF events reproduce well the experimental data, with values between 0.5 and 0.7.
We also observe a systematic small underestimation of the mean values of $R_{E}$ in the calculations, with respect to the experimental data.
As there is a difference in the definition of $M_{tot}$ in this work (where pre-equilibrium particles are excluded) and in~\cite{Lehaut}, 
we have checked if similar events are actually selected in the data and in the calculations.
We present in fig.~\ref{fig:DistriZMtotCut} a comparison of charge distributions for beam energies
above 39 MeV/nucleon. One can see that a rather good agreement is obtained. The absence of heavy fragment production in the calculations
is due to the lack statistics. The overestimation
of fragment  production in the Z range [10;30], at 39 and 45 MeV/nucleon, is a consequence of the $M_{tot}$ criterium. Being 
pre-equilibrium particles excluded in the theoretical analysis,  the corresponding $M_{tot}$ ranges are compressed
by about 30\% with respect to the experimental ones. This induces a loss of sensitivity 
in the selection and larger impact parameters may be kept in the selected calculated events, with respect to the data. 
The latter  contribute to the charge distribution mainly by fission of the quasi-projectile, leading to fragment production in the Z range [10;30] and to 
a small decrease of the $R_{E}$ value.
Nevertheless, in the high beam energy region, the  $M_{tot}$ selection gives similar sets of events in the calculations and in the data (see the right
panel of Fig.7) and general trends are not affected by the difference in the definition of $M_{tot}$.
Looking at the Fig.~\ref{fig:Stopping}.c (violet line), the simulation results appear shifted to the right, with respect to the data. 
A similar shift has been observed for the multifragmentation threshold. 
In fact, the results obtained for $R_E$ in the simulations (a decrease followed by a mild increase)
can be associated with the transition from statistical sequential decay of equilibrated sources, 
dominating at the lowest energies over the whole range of considered impact parameters,
to full multifragmentation of central collisions, observed at the highest beam energies.
The same mechanism could be present in the data, but with a lower ``transition'' beam energy (around 20 MeV/nucleon).    
 
Finally, we stress that, according to our simulations, the behavior obtained for
$R_E$ (Fig.~\ref{fig:Stopping}.c, violet line) comes essentially from the fact that 
the event selection based on the $M_{tot}$ cut-off induces some impact parameter mixing. 
Indeed, looking at the results corresponding to b=0.5 fm,
we observe full stopping in central collisions at Fermi energies,
in the beam energy range considered. This indicates that a full stopping scenario
could be compatible with the experimental results, though it cannot be taken 
as a conclusive proof of the reaction dynamics.
Thus the centrality selection 
in experimental data has to be carefully considered before proper conclusions can be drawn.

\section{Conclusion}

The issue of the link between the compression-expansion cycle, the collective radial expansion and 
the multifragmentation pattern observed in central collisions at Fermi energies
has been addressed comparing the predictions of the semi-classical SMF transport model
to INDRA multifragmentation data.

The simulations allow one to extract
the maximum radial expansion energy reached in reactions at 
beam energies in the Fermi domain. This can be used as a reference in flow estimation analyses 
based on statistical models. 
Looking at the time evolution of fragment formation, we show that
for all considered energies above 32 MeV/nucleon, density fluctuations occur and lead to pre-fragment 
formation already at t=100 fm/c. These nascent partitions have then different stories depending on the strength
of the velocity fields. For energies below 39 MeV/nucleon, the nascent partitions do not survive, leading 
to evaporation residue or fission fragments in the exit channel, while for higher energies, one observes 
multifragmentation events. Looking at the properties 
of these events, we obtain a good reproduction of data as far as the charge partitions are concerned.
Concerning kinematical properties, we observe and quantify the underestimation of velocities which suggests a 
weakening of initial radial expansion along the fragment formation process, 
due to the lack of thermal pressure effects, which is a typical drawback of semi-classical models.

We have also investigated the issue of stopping power in nuclear reactions and performed the same 
analysis proposed in~\cite{Lehaut}. We observe that for very central collisions, b=0.5 fm full stopping is achieved
at all energies. However, if a selection of centrality is done as in the experimental data, 
we recover the experimental results, which exhibit a lesser degree of stopping. 
We then pointed out the essential coherence needed
in the comparison between data and microscopic simulations in order to give unbiased indications on the microscopic 
ingredients of the considered models.
Within our approach a good reproduction of the experimental data, at beam energies above the multifragmentation
threshold,  is obtained by employing
the free (angle and energy dependent) NN cross section and a soft EOS (compressibility K = 200 MeV). 

Following the definition given in experimental studies,
we find that the evolution of the $R_E$ ratio with the beam energy
can be associated
with a change in the fragmentation mechanim, from statistical decay to prompt
multifragmentation. At the highest beam energies central events, where large fragment multiplicities are observed, 
start to dominate the selection experimentally considered (based on particle multiplicity cuts), thus leading to the small increase 
observed for the stopping power.
The simulation results appear shifted with respect to the experimental data, in analogy with the shift to higher energies observed,
in the model, for the multifragmentation threshold.

We conclude by mentioning that, in the framework of mean-fields approaches, 
new methods to improve the treatment of fluctuations, in the isovector channel \cite{Colonna_PRL} and in full phase space \cite{BLOB},
have recently been introduced. A more effective fluctuating term is expected to lead to a faster fragmentation process, thus lowering the fragmentation
threshold and enhancing fragment velocities. Further studies, based on the comparison with experimental data, are in progress.   

\section{Acknowledgments}

The authors thank the INDRA Collaboration for providing them the high quality data presented in this work.
They also thank M.F. Rivet for fruitful discussions. E.B. acknowledges the IN2P3 Computing Centre for providing
huge amounts of CPU time and data storage for the calculations.
\newpage



\end{document}